%% file: templateArxiv.tex
\documentclass{article}

\usepackage{PRIMEarxiv}

\usepackage[utf8]{inputenc} 
\usepackage[T1]{fontenc}    
\usepackage{hyperref}       
\usepackage{url}            
\usepackage{booktabs}       
\usepackage{amsfonts}       
\usepackage{nicefrac}       
\usepackage{microtype}      
\usepackage{lipsum}
\usepackage{fancyhdr}
\usepackage{amsmath}
\usepackage{graphicx}       
\usepackage{subcaption} 
\graphicspath{{media/}}     

\title{\textit{PINNslope}: seismic data interpolation and local slope estimation with physics informed neural networks
}

\author{
  Francesco Brandolin, Matteo Ravasi, Tariq Alkhalifah \\
  King Abdullah University of Science and Technology (KAUST) \\
  Thuwal, Kingdom of Saudi Arabia\\
  \texttt{\{francesco.brandolin, matteo.ravasi, tariq.alkhalifah\}@kaust.edu.sa} \\
}

\begin{document}
\maketitle


\input{Sections/abstract}

\input{Sections/Introduction}

\input{Sections/Theory}

\input{Sections/Results}

\input{Sections/Discussions}

\input{Sections/Conclusions}

\input{Sections/Acknowledgement}




\end{document}

%% file: Sections/abstract.tex
\begin{abstract}
Interpolation of aliased seismic data constitutes a key step in a seismic processing workflow to obtain high quality velocity models and seismic images. Building on the idea of describing seismic wavefields as a superposition of local plane waves, we propose to interpolate seismic data by utilizing a physics informed neural network (PINN).
In the proposed framework, two feed-forward neural networks are jointly trained using the local plane wave differential equation as well as the available data as two terms in the objective function: a primary network assisted by positional encoding is tasked with reconstructing the seismic data, whilst an auxiliary, smaller network estimates the associated local slopes. Results on synthetic and field data validate the effectiveness of the proposed method in handling aliased (coarsely sampled) data and data with large gaps. Our method compares favorably against a classic least-squares inversion approach regularized by the local plane-wave equation as well as a PINN-based approach with a single network and pre-computed local slopes. We find that introducing a second network to estimate the local slopes whilst at the same time interpolating the aliased data enhances the overall reconstruction capabilities and convergence behavior of the primary network. Moreover, an additional positional encoding layer embedded as the first layer of the wavefield network confers to the network the ability to converge faster improving the accuracy of the data term. 
\end{abstract}


%% file: Sections/Introduction.tex
\section{Introduction}

The idea of describing seismic data as a superposition of local plane-waves was introduced by Jon Claerbout \cite{Claerbout1992}. This elementary type of waves can be modelled by the plane-wave partial differential equation (plane-wave PDE), which is only parameterized by the local slope factor (also referred to as slowness or ray parameter). \cite{Claerbout1992} demonstrated how a plane-wave defined by its local slope can be annihilated within a given wavefield by means of a plane-wave partial differential operator. Leveraging this simple concept, he developed an approach for local-slope estimation using small moving windows across data, where a single slope at the center of the window is computed by linear least-squares. Later, \cite{Fomel2002} proposed the plane-wave destruction filters (PWD), a global approach for local slope estimation that requires the solution of a non-linear system of equations, therefore removing the need for windowing the data. Several techniques have been developed during the years that utilize the plane-wave approximation in seismic processing with applications ranging from denoising \cite{Canales1984}, trace interpolation, detection of local discontinuities \cite{Fomel2002}, velocity-independent imaging \cite{Fomel2007} and as a regularization in seismic inverse problems \cite{Fomel2006}. In this work we build upon the concept introduced by Claerbout, utilizing neural networks informed by the local plane-wave equation to simultaneously interpolate seismic data and estimate the local slopes of the associated events.\\
The idea of utilizing neural networks for estimating local slopes and interpolating seismic data has been recently explored in a number of publications.  \cite{Wu2019} proposed an approach based on supervised learning where a convolutional network (CNN) is trained to simultaneously perform three seismic image processing tasks: namely, detecting faults, enhancing the seismic image, and estimating structural orientation. \cite{Huang2021} developed a specific approach for slope estimation based on a hybrid architecture based on convolution and fully connected layers. \cite{Zu2022} estimated the slope from severely noisy seismic data (where instead the usual PWD technique fails) training a convolution-deconvolution network. Even if the results of all the proposed methodologies achieve accurate slope estimations, they all frame the slope estimation problem as a supervised learning task. This is a major drawback in seismic applications where a dataset of training samples and labels is needed; however, this is difficult or expensive to obtain (or generate). An unsupervised example of local slope estimation has been proposed by \cite{Bahia2022}. Building upon the deep image prior framework (initially developed by \cite{Ulyanov2020}, and consisting of a randomly initialized neural network used as prior to solve various types of inverse problems), they parameterize the unknown slope field by a CNN and optimize the network weights and biases by minimizing the plane-wave PDE residual, obtaining results similar to those of the PWD algorithm.\\
As far as seismic data reconstruction is concerned, \cite{mandelli2019} proposed a supervised learning approach based on patches, to denoise and interpolate regularly and irregularly subsampled shot gathers up to 50\% of missing traces, achieving good results with both synthetic and field examples. \cite{Garg2019} reconstructed aliased seismic data posing it as a super-resolution problem, using a CNN trained with a custom loss function that computes the errors both in space-time and frequency-wavenumber domains. \cite{Wang2019} also utilized a CNN trained with synthetic data generated through the finite-difference method to reconstruct aliased seismic data in the time-space domain. These methodologies heavily depend on the training data to perform the reconstruction task; however, creating a suitable training dataset that contains all the features of the seismic field dataset to be interpolated is often a challenging task. As a consequence, if the underlying distribution of the training and test are different, the result will be biased towards the features contained in the training set distribution. \cite{Fang2021} utilized an approach that combines deep neural networks and the classic prediction-error filter (PEF) to learn the relationship between the subsampled data and the PEFs, using later the learned filter for the recovery of the missing seismic traces. To avoid limitations arising from the need for training samples, \cite{Liu2021}, \cite{Kong2022}, and \cite{Min2023} utilized the deep-priors concept \cite{Ulyanov2020} that captures priors, based on the particular structure of the CNN, avoiding the need for any training data. Usually, the deep image prior algorithm requires a high number of epochs to converge, and transfer learning is often applied to make this method competitive in terms of performance. In this manuscript, we propose an approach based on physics informed neural networks (PINNs - \cite{Raissi2019}) that is fully unsupervised and that parametrizes both seismic data and the unknown local slope field with two separate neural networks, which are jointly trained.\\
PINNs have recently emerged as a novel computational paradigm in the field of scientific machine learning and have been shown to be effective in representing solutions of partial differential equations (PDEs). The PINNs framework can be utilized to solve both forward and inverse problems and has been successfully applied in various domains of computational physics (\cite{Chen2020}, \cite{MAO2020}, \cite{Cai2021}). In the context of exploration seismology, PINNs have been utilized with different PDEs, for modelling wave propagation in time domain using the wave equation (\cite{Moseley2020}, \cite{Ren2022}) and to model wavefields in frequency domain applying the Helmholtz equation and the vertical transversely isotropic wave equation (\cite{ALKHALIFAH2021}, \cite{CSong2021} and \cite{Konuk2021}). PINNs have also been applied to the eikonal equation, especially to overcome some of the limitations of conventional seismic tomography methods. For example, while classical traveltime tomography uses a generic non-physics based smoothing regularization to compensate for the ill-posedness of the problem (limiting the resolution of the inverted velocity model), with PINNs the residual of the eikonal equation is utilized, embedding the actual wave propagation physics as a regularizer (\cite{Waheed2021}, \cite{Smith2021} and \cite{Chen2022}).
PINNs are a special class of feed-forward neural networks that estimate the solution of a PDE, which governs the physics of the system, restricting the space of possible solutions in favor of physically plausible ones. In the standard scenario of parameter estimation problems, where the loss function includes a data fitting term (as in our implementation), the PDE term acts as a soft constraint to the network optimization problem. Specifically, PINNs are generic function approximators promoting reconstructions that are naturally similar to the available data. With this advantage, we can obtain an algorithm that leverages on the interpolation capabilities of feed-forward neural networks, but whose solution is heavily dependent on the physics of the problem. Additionally, PINNs are mesh-free given that once trained, they can be evaluated at any point within the training domain. Moreover, compared to traditional numerical approaches that rely on the discretization of derivatives involved in a PDE, PINNs learn a direct mapping from spatial coordinates to wavefield amplitudes (or any other physical quantity), removing the need for finite difference approximations, and relying on functional derivatives (automatic differentiation), which are more stable and accurate.\\ 
Once a PINN is trained, the solution can be rapidly evaluated at any point in the computational domain defined by the input grid points. However, if the setting of the problem changes (e.g., different geometry or values of the medium under consideration, different position of the source functions, or a different dataset, in an inversion scenario), the network has to be retrained, incurring additional computational cost. Furthermore, the choice of the architecture capacity and hyperparameters is critical for PINNs and some variations to the original architecture (number of layers, type of activation function) can be further required when applying it to different datasets or model settings.
Moreover, PINNs are parametrized by a simple multilayer perceptron (MLP) architecture and therefore are affected by the well known low frequency bias of neural networks \cite{rahaman2019}. The property of this type of architecture makes PINNs suffer from frequency dependent learning speed where lower frequencies are being fitted first during training.\\
A deep learning framework named coordinate based learning has emerged in the literature (\cite{Mildenhall2021}, \cite{Sun2021} ). This approach aims to solve imaging inverse problems by utilizing an MLP to learn a continuous mapping from the measurement coordinates to the corresponding sensor responses. However, compared to PINNs, this approach does not utilize any physics laws to constrain the network solution. Moreover, similar to PINNs it has been shown to perform poorly at representing the high-frequency component of the signals. To overcome this limitation, \cite{Mildenhall2021} proposed to map the input coordinates to a higher dimensional space through $positional$ $encoding$ (or Fourier Features encoding) before feeding them to the network. Successful implementations of $positional$ $encoding$ in seismic applications can be found in the works of \cite{Huang} and \cite{Goyes2022}, which proposes an anisotropic version of positional encoding in a coordinate based learning seismic interpolation application, and in \cite{Song2023} that simulate seismic multi-frequency wavefields using a PINN with embedded Fourier features. In our implementation, we also utilize positional encoding to ensure that the network is capable of reconstructing multi-scale, oscillating signals like those encountered in seismic data.\\
To support our claims, we present numerical examples focused on two of the most challenging tasks in seismic interpolation: namely, interpolation of regularly subsampled data (beyond aliasing) and data with large gaps.  We first evaluate the performance of the proposed approach on two simple synthetic data examples, comparing the results obtained by the proposed PINNs approach with a simple plane-wave regularized least-square inversion (PWLS) and against a previous version of our framework called PWD-PINN \cite{Brandolin2022}. We then consider a field data example where given the challenge associated with the recovery of the higher frequency content and higher complexity of the recorded signals, we embedded an anisotropic positional encoding layer to overcome the low frequency bias of these types of architectures.
The proposed framework, which we refer to as PINNslope, not only allows to interpolate seismic data, but it can also be used to estimate slopes from fully sampled data of quality comparable to those of the PWD filters. In the context of sparsely sampled data, the procedure is advantageous because we can directly estimate the slopes during the interpolation of the recorded wavefield (i.e. using the aliased data), without the need for low-pass filtering of the data to obtain an alias-free version on which to perform a reliable slope estimation, as needed in our previous work \cite{Brandolin2022}. The inversion result of the slope network turns out to be smooth, but a more accurate version than the one estimated from the low frequency data by means of PWD filters. 
To summarize, our main contributions comprise:
\begin{enumerate}
    \item A novel machine learning framework for slope assisted seismic data interpolation. 
    \item An innovative procedure for local slope attribute estimation by the mean of physics informed neural networks.
    \item Successful application of the framework on field data.
\end{enumerate}

The paper is organized as follows. First, we present the theoretical background of our methodology. We then describe the network architecture, focusing on some of the key implementation details needed to achieve a stable training process. Finally, our method is applied to a range of synthetic and field data. 

%% file: Sections/Theory.tex
\section{Theory}

\subsection{Modelling operator}
The objective of this paper is to formulate the problem of seismic data interpolation within the framework of physics informed neural networks.
A dataset of seismic traces at desired locations is defined as
  $\textbf{u}=\left[\begin{smallmatrix}
  \textbf{u}_{1} \\
  \textbf{u}_{2} \\
  \vdots \\
  \textbf{u}_{N_{u}}
\end{smallmatrix}\right]$, where $\textbf{u}_{1}, \textbf{u}_{2}, ..., \textbf{u}_{N_{u}}$ are column vectors of dimension $ N_{t} \times 1$ (with $N_{t}$ being the number of time samples) vertically stacked together and $N_{u}$ is the number of total traces in the dataset. The basic mathematical model used to obtain a decimated version of the original seismic data is a restriction operator $\textbf{R}$ of dimension ($N_{s}N_{t}\times N_{u}N_{t}$), defined such that it samples the vector $\textbf{u}$ at the desired locations, removing the traces that at a later stage we wish to interpolate (with $N_{s}$ being the number of the traces that we keep in the dataset). In matrix notation, the operation of subsampling a gather of traces can be written as:
\begin{equation}
     \textbf{d} = \textbf{R}\textbf{u}
\end{equation}
where $\textbf{d}$ is the subsampled data that is either missing traces at a regular interval to simulate spatial aliasing, or a large number of consecutive traces to simulate a gap in the acquisition geometry.

\subsection{Slope estimation with plane-wave destructors}

The physical model used to express seismic data as local plane-waves is represented by the local plane-wave differential equation:
\begin{equation}\label{eq:1}
    \frac{\partial u(t,x)} {\partial x}+\sigma(t,x)\frac{\partial u(t,x)} {\partial t} = r(t,x)\approx0,
\end{equation}
where $u(t,x)$ is the pressure wavefield, $r(t,x)$ is the PDE residual, and the parameter $\sigma(t,x)$ is the local slope (or wavenumber) with units equal to the inverse of the velocity of propagation.
An analytical expression exists for the solution of equation \ref{eq:1} in case of a constant slope, which is simply represented by a plane-wave
\begin{equation}\label{eq:2}
    u(t,x)= f(t-\sigma x),
\end{equation}
where $f(t)$ is an arbitrary waveform at $x=0$. We can see that the left hand side of equation \ref{eq:1} decreases as the observation $u(t,x)$ matches the wave displacement $u(t-\sigma x)$ \cite{Claerbout2014}.\\
In our work, we are interested in computing a slope varying both in time and space, for which no analytical solution exists. Hence, Claerbout (1992) casts the dip estimation process as a linear least-squares problem, through an operation named plane-wave destruction. In this approach, the curvature of the events is linearly approximated by computing the slope in a small window of the entire data. The slope is estimated through equation \ref{eq:1}, minimizing the quadratic residual:
\begin{equation}
    Q(\sigma) = (\textbf{u}_{x}+\sigma\textbf{u}_{t})\cdot(\textbf{u}_{x}+\sigma\textbf{u}_{t}),
\end{equation}
where $\textbf{u}_{x}$ and $\textbf{u}_{t}$ are respectively defined as the spatial and temporal derivatives of the wavefield \textbf{u}. Setting the derivative of $Q(\sigma)$ to zero, we can find its minimum as:
\begin{equation}
    \sigma = -\frac{\textbf{u}_{x}\cdot\textbf{u}_{t}}{\textbf{u}_{t}\cdot\textbf{u}_{t}}.
\end{equation}

Fomel (2002), on the other hand, frames the slope estimation as a non-linear least-squares problem computing the slope globally on the entire data by means of plane-wave destruction filters.
Given the vector $\textbf{u}$ (seismic traces vertically stacked together), the destruction operator predicts each trace from the previous one by shifting the observed trace along the dominant local slopes of the seismic data and subtracts the prediction from the original one \cite{Fomel2010}. 
The phase-shift operation on the traces is approximated by an all-pass digital filter (or prediction filter in 2D). The filter coefficients are determined by fitting the filter frequency response (at low frequencies) to the response of the phase-shift operator. In this implementation, the slope ($\sigma$) enters in the filter coefficients in a non-linear way. To characterize and describe the entire seismic gather, the prediction of several plane waves is therefore needed (and not only one). This is achieved by cascading various filters of the above mentioned form.
The filters are ultimately applied to the data $\textbf{u}$ as a convolutional operator $\textbf{D}(\sigma)$. In matrix notation, the slope estimation problem can be written as
\begin{equation}\label{eq:4}
    \textbf{D}(\sigma)\textbf{u}=\textbf{r},
\end{equation}
where $\textbf{r}$ is the residual. This non-linear least-squares problem is solved via Gauss-Newton iterations, which implies solving
\begin{equation}\label{eq:4}
    \textbf{D}'(\sigma_{0})\Delta\sigma\textbf{u}+\textbf{D}(\sigma_{0})\textbf{v}=\textbf{r},
\end{equation}
where $\textbf{D}'(\sigma_{0})$ is the derivative of the filter coefficients $\textbf{D}(\sigma)$ with respect to $\sigma$ (which in the case of PWD is unitless). The minimization problem is solved for the update $\Delta\sigma$, which is repeatedly added (at every iteration) to an initial guess $\sigma_{0}$. The problem can be regularized by adding an appropriate penalty term that avoids oscillatory solutions of the slope attribute.

\subsection{Plane-wave regularized least-squares interpolation}

In this section, we describe a regularized least-squares approach to take into account pre-computed slopes whilst interpolating seismic data (i.e., restoring missing traces). This method will be later used as a benchmark for our PINNs approach.
The inverse problem is cast as follows: finding the vector $\textbf{u}$ (i.e., the full set of traces) that minimizes the Euclidean distance between the subsampled data $\textbf{d}$ and the estimated subsampled data $\textbf{R}\textbf{u}$, whilst at the same time satisfying the plane-wave differential equation with pre-computed slopes. The objective function is formally defined as
\begin{equation}
    f(\textbf{u}) = \|\textbf{d}-\textbf{R}\textbf{u}\|_{2}^{2}+\epsilon_{r}\|\textbf{u}_{x}+\boldsymbol\Sigma\textbf{u}_{t}\|_{2}^{2}
\end{equation}
where $\textbf{u}_{x}$ and $\textbf{u}_{t}$ are respectively defined as the spatial and temporal derivatives of
the data $\textbf{u}$, $\boldsymbol\Sigma$ is a diagonal matrix that applies element-wise multiplication of the pre-computed local slope and $\epsilon_{r}$ is a weight to control the contribution of the PDE in the solution. The data term of the objective function aims at accurately reproducing the available traces from the estimated full shot-gather $\textbf{u}$ subsampled by the restriction operator $\textbf{R}$. This means that the interpolation operation between the available traces is effectively performed by the local plane-wave regularization term. In other words, the regularization term has the role of filling the gaps between the subsampled traces, spraying the information available from two neighboring traces along the direction of the provided local slope field.

\subsection{Physics Informed Neural Networks}
\begin{figure}[h]
    \centering
    \includegraphics[scale=0.55]{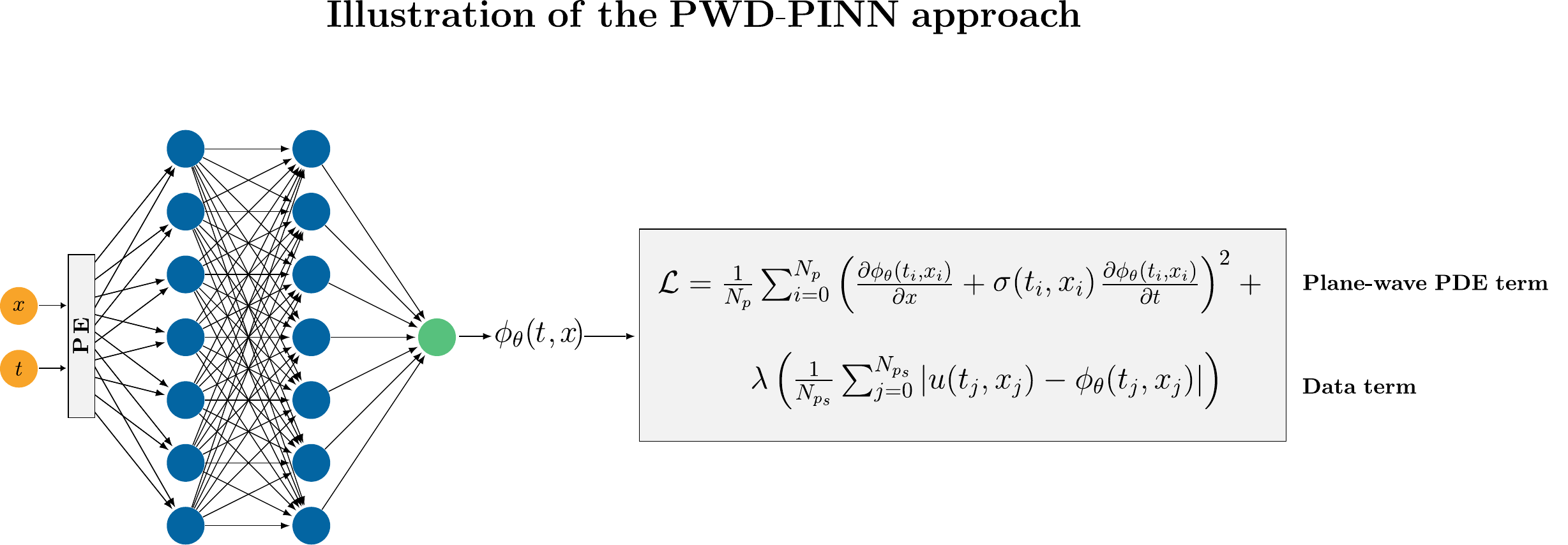}
    \caption{A diagram explaining the PWD-PINN algorithm. The network is trained while maintaining the slope array fixed during training.}
    \label{fig:PWDPINN}
\end{figure}

In this section, we aim to show that starting from the knowledge of a slope field, estimated via PWD filters or any other algorithm, the problem of seismic interpolation can be formulated within the PINNs framework.
PINNs have been designed to blend the universal function approximator capabilities of neural networks \cite{Hornik1989} with a physical constraint given by a PDE, which describes the physical system under study.
In our specific case, the PDE that we seek to satisfy is the local plane-wave differential equation (equation \ref{eq:1}).\\
A neural network $\phi_{\theta}(t,x)$ is designed to approximate the function $u(t,x)$, where $\theta$ refers to the weights (and biases) to be optimized and the coordinates pair ($t,x$) represents the input to the network. The network predicts the recorded wavefield $u(t,x)$ at the corresponding location in the time-space domain of interest.
A remarkable convenience of PINNs is that in contrast to traditional numerical methods, they do not require a discretization of the computational domain.
The partial derivatives of the underlying PDEs are computed by means of automatic differentiation (AD), which is a general and efficient way to compute derivatives based on the chain rule. AD is usually implemented in neural networks training to compute the derivatives of the loss function with respect to the parameters of the network. However, AD can be more broadly applied to every computational program that performs simple arithmetic operations and calculates elementary functions (linear transformations and non-linear activation functions in the case of neural networks) by keeping track of the operations dependencies via a computational graph and successively computing their derivatives using the chain rule.
The PINN framework is trained in an unsupervised manner, using a loss function that includes both the local plane-wave differential equation and a set of $N_{p_{s}}$ (number of grid points corresponding to the available traces in the subsampled gather) boundary conditions representing the subsampled traces of the gather in the computational grid:\\
\begin{equation}\label{eq:5}
    \begin{aligned}
    \mathcal{L} = \frac{1}{N_{p}}\sum_{i=0}^{N_{p}}\left(\frac{\partial \phi_{\theta}(t_{i},x_{i})} {\partial x}+\sigma(t_{i},x_{i})\frac{\partial \phi_{\theta}(t_{i},x_{i})} {\partial t}\right)^{2} + \lambda\left(\frac{1}{N_{p_{s}}}\sum_{j=0}^{N_{p_{s}}}|u(t_{j},x_{j})-\phi_{\theta}(t_{j},x_{j})|\right),
    \end{aligned}
\end{equation}

where $(t_{i},x_{i})$ are points randomly sampled from the input coordinate grid points (with $N_{p}$ being the total number of grid points in the computational grid). $u(t_{j},x_{j})$ are the known traces at points indexed by $j$. This set of grid points corresponds to the available traces, which we are trying to reproduce as closely as possible. The fidelity of the fitting is directly dependent on the scalar weight $\lambda$, whose value is currently set by trial and error. The weight $\lambda$ is applied to the data term because we find it easier to directly tune the influence of the data term in the overall solution (instead of weighting the PDE term). In practice, by looking at the different contributions of the loss function one can determine which of the terms has the major contribution in the solution. If the solution is not the aimed one, the parameter $\lambda$ can be adjusted consequently.\\
In this first approach, named PWD-PINN (Figure \ref{fig:PWDPINN}), the slope $\sigma(t_{i},x_{i})$ is pre-computed, for example, estimated utilizing PWD filters prior to the training process of the network. The slope field remains fixed during training. 

Differently from the plane-wave regularized least-squares (PWLS) method described in the previous section, which directly optimizes the wavefield, PWD-PINN optimizes (or learns) the network parameters. Once all the weights and biases have been learned, we feed again the input grid points to PWD-PINN to output the interpolated seismic gather. Hence, another advantage of utilizing a coordinate based neural network (like PINNs) resides in the fact that after training the interpolated data is stored inside the network in the form of the network parameters, making it useful for subsequent applications.

\subsection{Simultaneous data interpolation and slope estimation}

In this section, we introduce the slope estimation framework using physics informed neural networks, named PINNslope. 
We propose to estimate the local slope while at the same time interpolating the aliased data (or any other type of interpolation task). 
Specifically, we simultaneously train two neural networks to predict the data and the local slope that jointly satisfy the plane-wave PDE. This approach bears similarity with previous works by \cite{Waheed2021} and \cite{taufik2022}, in the context of traveltime tomography.

\begin{figure}[h]
    \centering
    \includegraphics[scale=0.55]{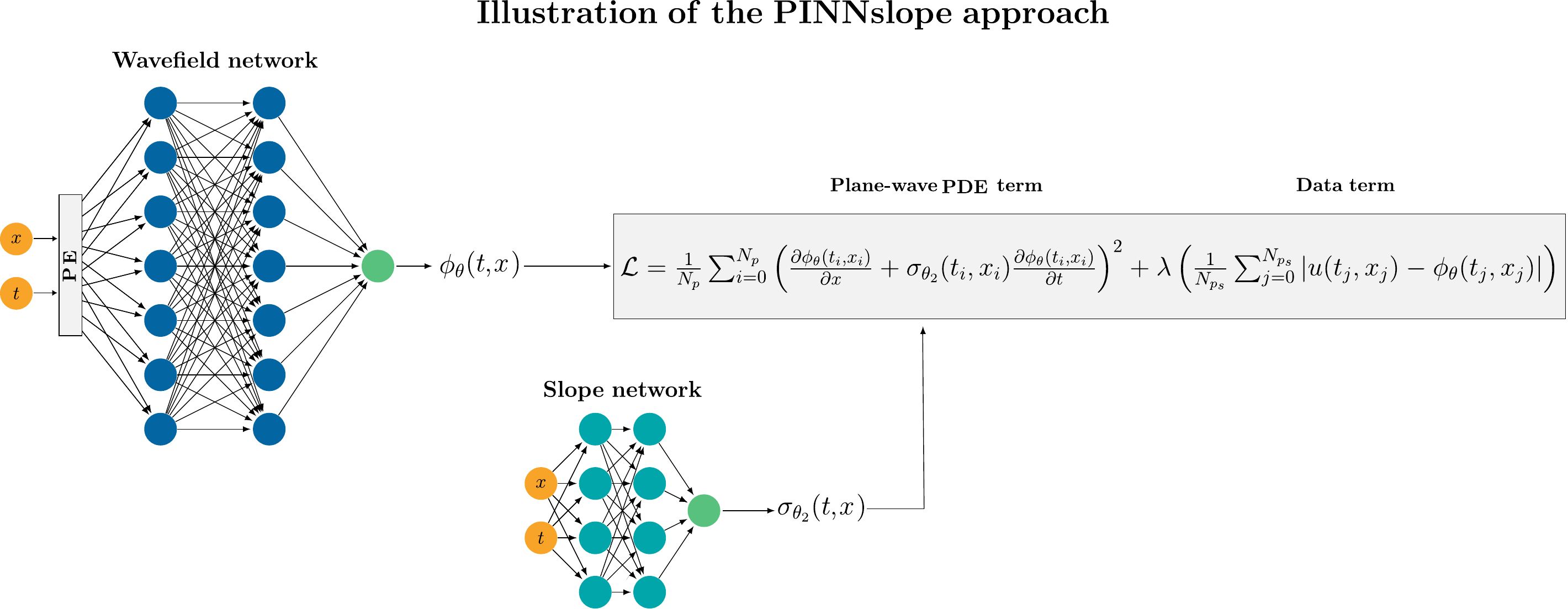}
    \caption{Double network scheme for joint wavefield and slope estimation in the same training procedure.}
    \label{fig: Double - Net}
\end{figure}

As shown in the diagram of fig.\ref{fig: Double - Net}, both networks have fully-connected architectures and utilize $Tanh(\cdot)$ activation functions; moreover, a positional encoding layer is added to the wavefield network to enhance the network capabilities to reconstruct multi-scale signals (i.e., signal exhibiting both low and high frequency components) such as our seismic traces. The two networks also differ in the number and size of the layers: the wavefield network aims at reconstructing the shot gather data, and therefore it requires a much larger number of degrees of freedom to fit the complexity of the seismic signals (i.e., the available seismic traces $u(t_{j}, x_{j} )$); rather for the slope, as we aim to obtain a smooth solution, this can be achieved by using a more compact architecture. After computing the loss function, both networks are simultaneously updated. Two separate ADAM optimizers \cite{kingma2017} are utilized to allow two different learning rate values if necessary.

\subsection{Positional Encoding}
During the numerical experiments, both frameworks initially struggled to fit signals with high frequency content. In our previous experiments, the low frequency bias of neural networks \cite{rahaman2019} was addressed with frequency upscaling by the mean of \textit{neuron splitting} \cite{Huang2022} and with locally adaptive activation function \cite{Jagtap2020}. Here, the low frequency bias of multi-layer perceptrons (MLPs) is tackled by including  \textit{positional encoding} to the spatial coordinates \cite{Mildenhall2021}.\\
Differently from the classical Transformer approach to positional encoding, where its usage is needed to track token positions, in our application positional encoding is used to simply map the input coordinate grid into a higher dimensional space, which allows for a better fitting of high frequency signals. The approach implemented in this work resembles the one previously presented in \cite{Sun2021} and referred to as \textit{Fourier feature mapping}, where the authors utilized a linear sampling in the Fourier space that enables a large amount of frequency components in the low frequency regions. The modification of \textit{Fourier feature mapping} that we propose is fundamental to the success of our implementation, as other forms of encodings were introducing small artifacts in the reconstruction, or they were not able to accurately interpolate the shot gather following the slope information. Additionally, the modified encoding function follows the idea proposed by \cite{Goyes2022} of an anisotropic version of positional encoding, justified by the idea that seismic data components present different features and should not be equally encoded. Given a coordinate matrix defined as $\textbf{C}=[\textbf{x},\textbf{t}]$ with $\textbf{x}$, $\textbf{t}$ being two column vectors that respectively denote the coordinates (grid points) in space and time, the \textit{modified} \textit{Linear Fourier Features} (modified LFF) encoding in the space dimension is calculated as
\begin{equation}\label{mod FF enc}
\begin{centering}
     \gamma_{N_{X}}(\textbf{x})=\Bigl[cos \left( a\pi \textbf{x}\textbf{k}_{x}^{T} \right), sin \left(a\pi \textbf{x}\textbf{k}_{x}^{T} \right) \Bigr],
\end{centering}
\end{equation}
where $a$ is a constant scale factor and $\textbf{k}_{x}$ is a column vector $\textbf{k}_{x}=[0,1,...,N_{X}-1]$, composed of an ordered set of evenly spaced elements. Here, $N_{X}$ is the total number of encoding frequencies in the space (offset) dimension. The formula for the encoding of the time coordinates $\gamma_{N_{T}}(\textbf{t})$ is the same as in equation \eqref{mod FF enc}, with the vector $\textbf{k}_{t}$ constituted by $N_{T}$ linearly spaced elements ($N_{T}$ is the total number of encoding frequencies in time). The encoded coordinates of all four functions are subsequently concatenated column-wise as
\begin{equation}
     \Gamma_{X,T}(\textbf{x},\textbf{t})= [ \gamma_{X}(\textbf{x}),  \gamma_{T}(\textbf{t})].
\end{equation}
The positional encoding operation is finally embedded inside the architecture of the data network as follows
\begin{equation}
     \textbf{x}^{[1]}_{out} = tanh(\textbf{W}^{[1]}\Gamma_{X,T}(\textbf{x},\textbf{t})+\textbf{b}^{[1]})
\end{equation}
where $tanh$ is the layer activation function, while $\textbf{W}^{[1]}$ and $\textbf{b}^{[1]}$ are the weights and biases of the first MLP layer and $ \textbf{x}^{[1]}_{out}$ is the output of the first MLP layer of the PINN architecture. The number of encoding frequencies $N_{X}$ and $N_{T}$ have been empirically determined by training our PINNs several times, picking at last pairs of values that returned the best interpolation result.

%% file: Sections/Results.tex
\section{Numerical experiments}
In this section, the proposed methodology is tested on synthetic and field data. For both the PWD-PINN and PINNslope approaches, a feed-forward neural network architecture with 4 layers and a $Tanh(\cdot)$ activation function was utilized. We set the number of neurons to be the same for all layers in all our experiments; however, this number can vary in different experiments as we specify it in the subsections. In both frameworks, the networks are trained in an unsupervised manner, passing as input an ensemble of $(x,t)$ points. The ensemble is passed to networks in batches of 1000 randomly sampled points. For every batch, the ensemble of the collocation points is concatenated to an array containing half of the points $(x_{j}, t_{j})$ associated with the available traces to be fitted.
All networks in every experiment are trained using ADAM optimizer, with a learning rate fixed at $10^{-3}$.
These parameters have been tuned running the network several times and assessing the reconstructed wavefields that produce the best results in terms of signal-to-noise ratio ($SNR = 10\log_{10}\left(\frac{\sum_{i=1}^{N_{p}}||d_{ref}||^{2}}{\sum_{i=1}^{N_{p}} ||d_{ref}-d_{out} ||^{2}}\right)$, where $d_{ref}$ is the reference data and the $d_{out}$ the network prediction) and mean squared error ($MSE=\sum_{i=1}^{N_{p}} ||d_{ref}-d_{out} ||^{2}$), subsequently they have been kept fixed throughout the study for all of the experiments.\\
All the experiments have been performed on Intel(R) Xeon(R) CPU E5-2680 v2 @ 2.80GHz equipped with a single NVIDIA GeForce RTX 3090 GPU.\\

\subsection{Synthetic data examples}

\subsubsection{Local slope estimation}
In this first example, we estimate the slope with the PWD algorithm and with the PINNslope framework, to compare their performance. The synthetic seismic image (Sigmoid model, \cite{Claerbout1992}) is assumed to be fully sampled and all the traces have been utilized in the training process.

\begin{figure}[h!]
    \centering
    \includegraphics[scale=0.4]{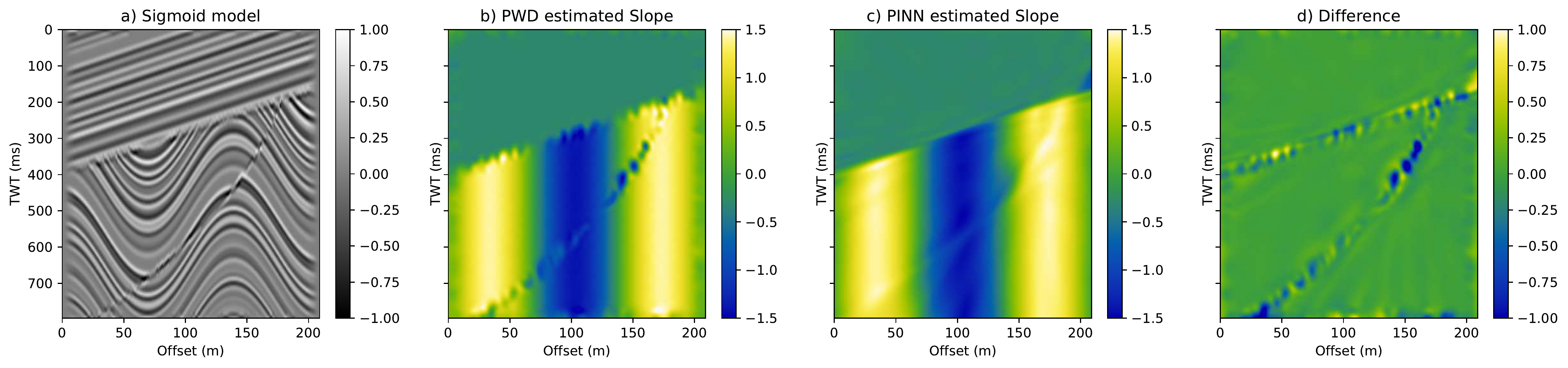}
    \caption{Slope comparison between PWD algorithm. a) Seismic image, b) local slope estimate from Plane-wave destruction filters 
    \cite{Fomel2002}, c) local slope estimate obtained with the PINNslope approach, d) difference between the PWD and PINNslope estimated slopes.}
    \label{Sigmoid}
\end{figure}
As shown in Figure \ref{Sigmoid}, the PINNslope framework can accurately estimate the local slope of complex subsurface geometries, and it results in a slightly smoother version with fewer artifacts near the major fault compared to the local slope estimated via the PWD algorithm.

\subsubsection{Interpolation beyond aliasing with local slope estimation}
\begin{figure}[h]
    \centering
    \includegraphics[width=1\textwidth]{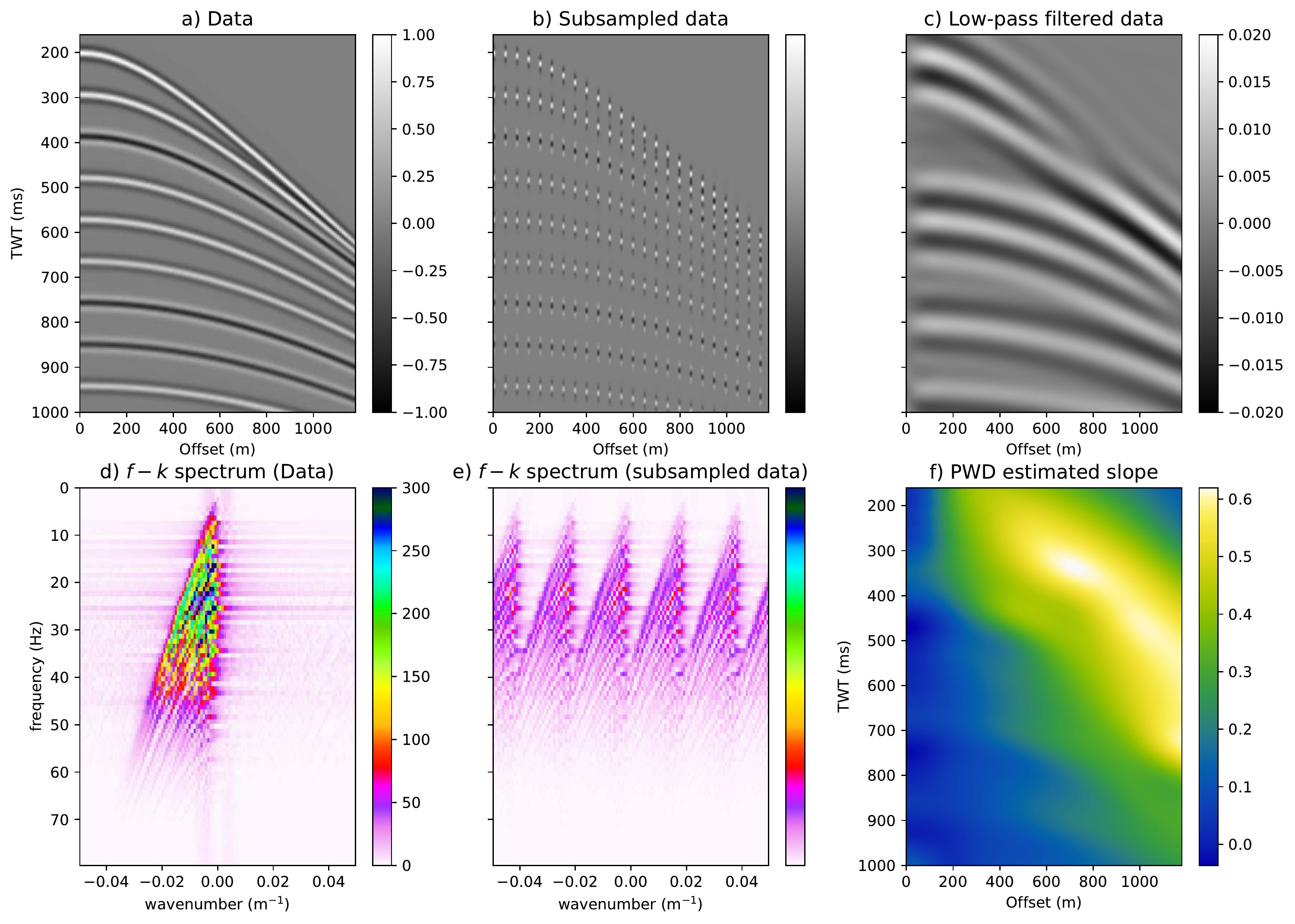}
    \caption{a) Original seismic data, b) seismic data with missing traces, c) low frequency data from which the PWD slope has been computed, d) $f-k$ spectrum of the data, e) $f-k$ spectrum of the subsampled data, f) PWD estimated slope from the low frequency data in Figure \ref{fig:introsynth}c.}
    \label{fig:introsynth}
\end{figure}

The goal of this second example is to reduce the spatial aliasing present in the recorded data by interpolating the missing traces. The synthetic data in Figure \ref{fig:introsynth}a have a trace spacing of 10 meters (with a time sampling interval of 0.004 seconds) and have been subsampled by a factor of 5 through the operator $\textbf{R}$, to obtain the aliased version in Figure \ref{fig:introsynth}b.

It is not possible to apply the PWD filter directly to estimate the slope from the subsampled data since this can lead to erroneous estimates, as the algorithm will estimate the aliased dips instead of the true ones of the fully sampled data.
To avoid this issue, the following pre-processing steps have been performed (similar to what is presented in \cite{Gan2015}):
\begin{enumerate}
    \item Apply $f-k$ filter to the spectrum of the aliased data.
    \item Inverse transform the filtered spectrum to get a low frequency alias-free version of the data.
    \item Apply the PWD filters algorithm to the low frequency data and estimate the slope.
    \item Utilize the PWD estimated slope from low frequency data inside the PINN loss function.
\end{enumerate}

The network capacity corresponds to 4 layers with 512 neurons each, with the number of encoding frequencies set to $N_{X}=8$ and $N_{T}=32$ for the $x$ and $t$ coordinates respectively. The network is trained using the loss function in equation \ref{eq:5} with the parameter $\lambda$ set to 1000 and $\sigma$ corresponding to the PWD estimated slope displayed in Figure \ref{fig:introsynth}c. 
\begin{figure}[h!]
    \centering
    \includegraphics[scale=0.40]{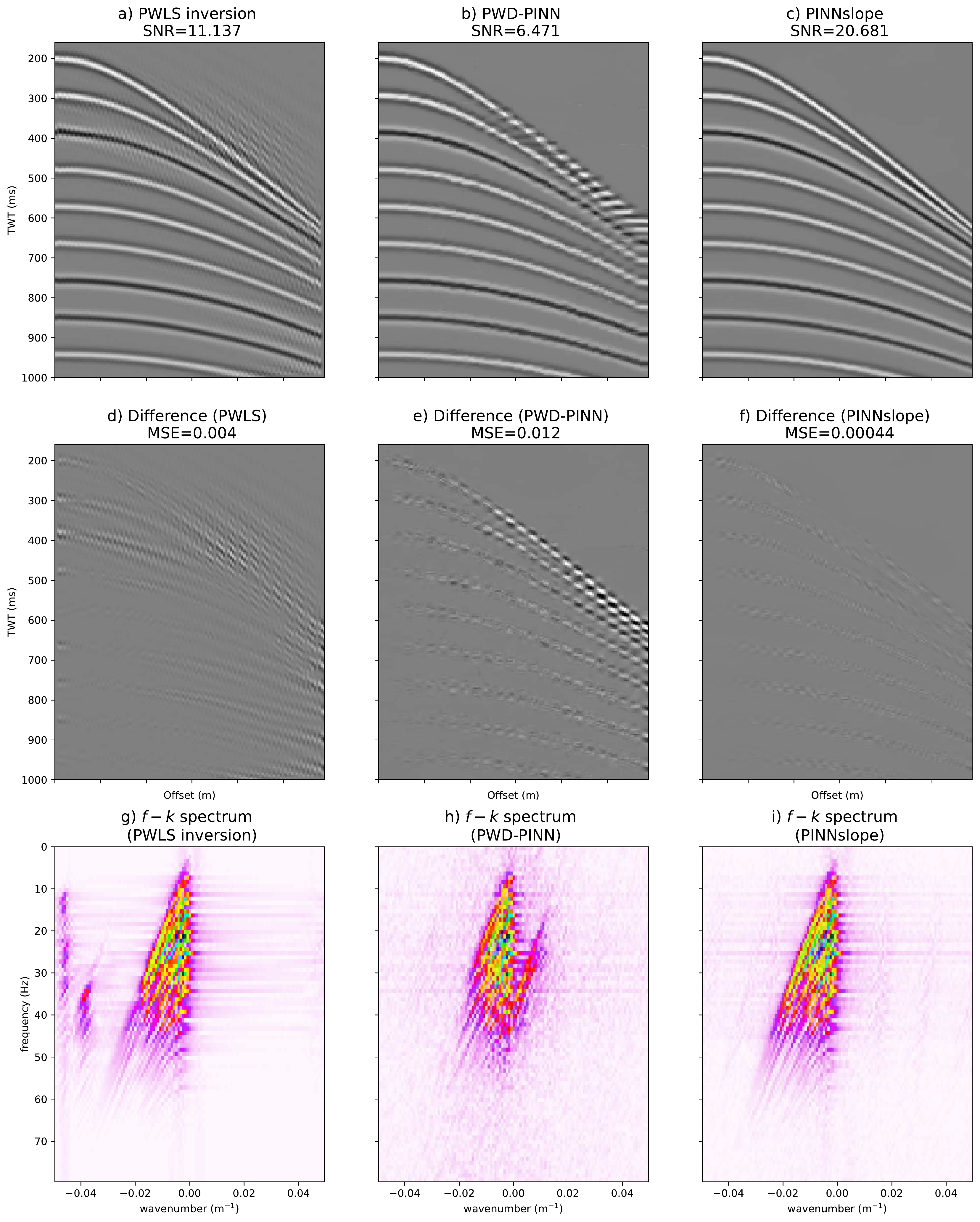}
    \caption{From a) to c): interpolation results of plane-wave regularized least-squares inversion, PWD-PINN and PINNslope. From d) to f) differences between the original data and PWLS inversion, PWD-PINN, and PINNslope. From g) to i) $f-k$ spectrum of PWLS inversion, PWD-PINN and PINNslope. The SNR and MSE values are provided on top of the panels.}
    \label{fig:synth_comparison}
\end{figure}
Figure \ref{fig:synth_comparison} compares the results obtained with the different approaches (the signal-to-noise ratio (SNR) and mean squared error (MSE) values of the differences are provided on the top of the figure panels). The output of the regularized least-squares inversion in Figure \ref{fig:synth_comparison}a demonstrates the importance of the plane-wave penalty term, which helps in filling the gap between the available traces following the correct overall geometry of the arrivals. Unfortunately, as soon as the reflections start bending their resolution decreases, worsening towards the far-offset. In this interpolation attempt, the sharp and definite seismic response that characterizes this simple synthetic data is slightly spread in a fuzzy pattern, a sign that the algorithm cannot properly restore the energy in the correct position. The difference with the original data in Figure \ref{fig:synth_comparison}d  shows some clear signal leakage, as well as some other artifacts. The achieved result is almost perfect where the arrivals are generally linear. To conclude, we however note that the PWLS inversion is almost instantaneous compared to the neural networks approach. The result displayed in Figure \ref{fig:synth_comparison}b requires in fact a runtime of approximately 26 minutes for 2000 epochs as shown in the plot of the loss curves in Figure \ref{fig:lossessynth}a.

\begin{figure}[h!]
    \centering
    \includegraphics[width=1\textwidth]{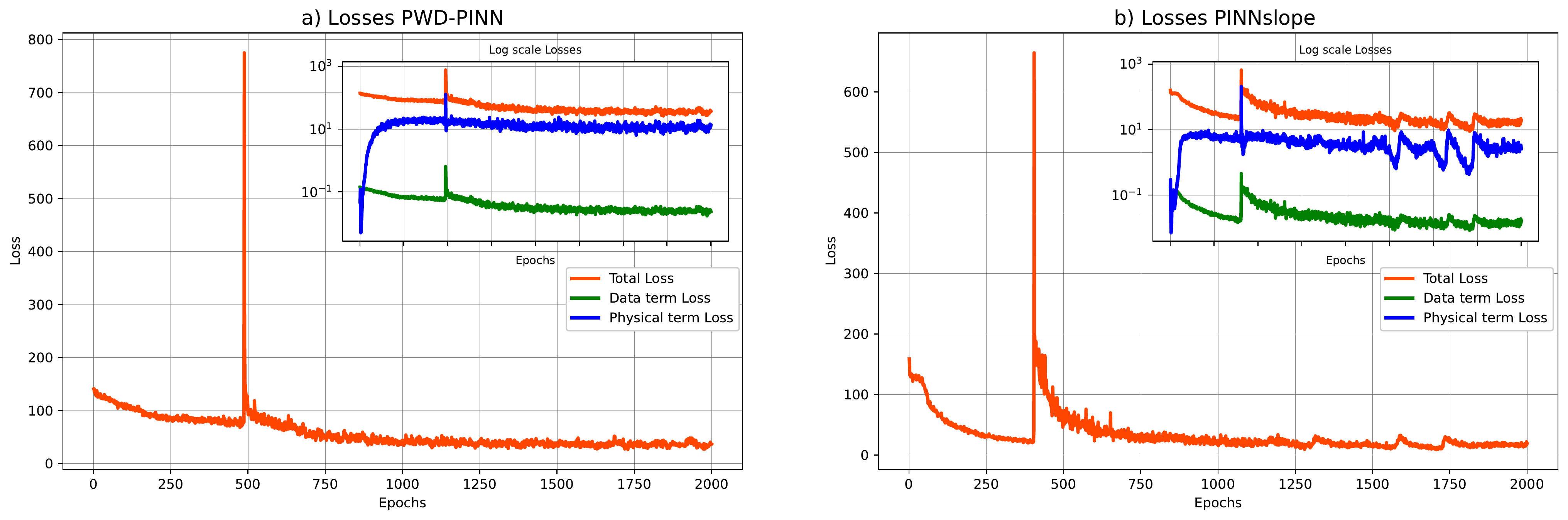}
    \caption{a) Loss of the PWD-PINN training. b) Loss resulting from the PINNslope training. In the small box inside the two plots it is shown the contribution of each term of the loss function in log-scale, that is: data-term (in green), physical-term (in blue), and total loss (in orange).}
    \label{fig:lossessynth}
\end{figure}

\begin{figure}[h]
    \centering
    \includegraphics[scale=0.4]{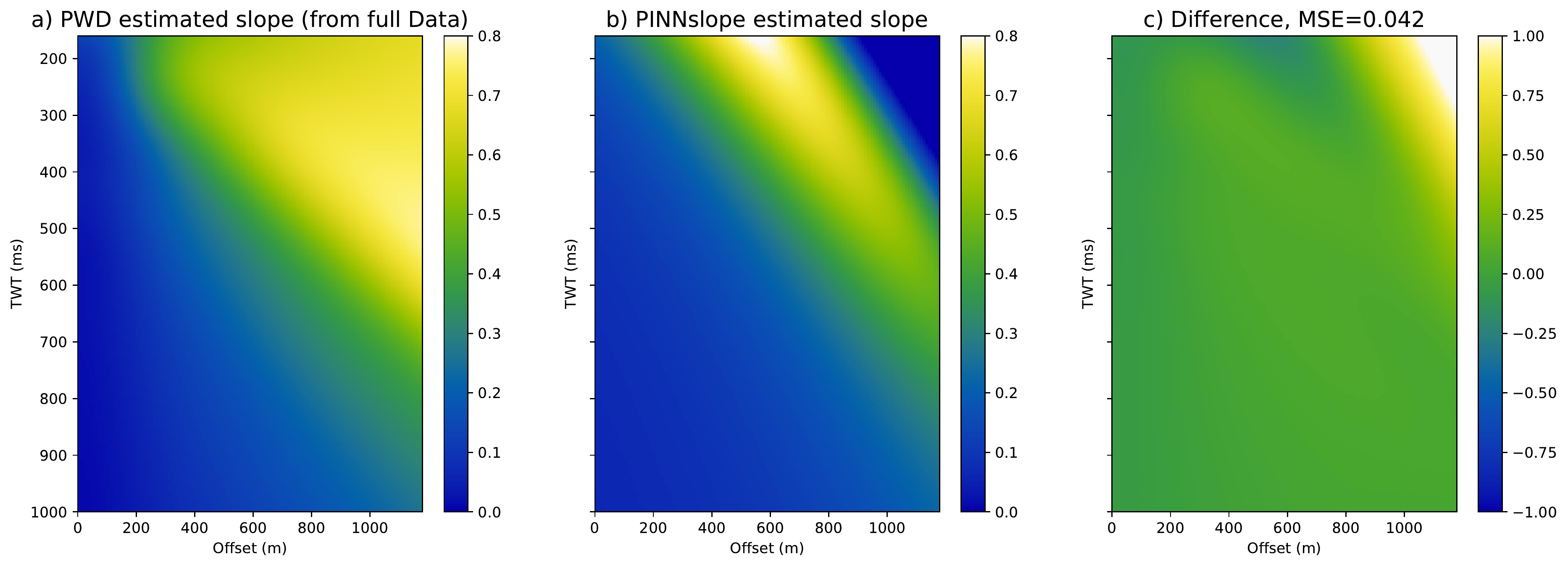}
    \caption{a) Accurate slope estimated with PWD algorithm from the data shown in Figure \ref{fig:introsynth}a, b) Slope estimate through the PINNslope framework while simultaneously interpolating the data in Figure \ref{fig:synth_comparison}c, c) difference between Figure \ref{fig:synthPINNslope}a and Figure \ref{fig:synthPINNslope}b.} 
    \label{fig:synthPINNslope}
\end{figure}

As in the previous result, the quality of the interpolation of the PWD-PINN algorithm decreases in the far-offset, although only for the first few reflections. This is a limitation of the algorithms that leverage on the PWD estimated slope, which is inaccurate at the far-offsets where the events are steeper; the slope estimated via PWD inherently contains errors because of the procedure through which it has been computed, but even more, it has been estimated from a low frequency version of the original data. Despite the poor interpolation of the above mentioned arrivals, all the others look adequately restored. Most of the energy is in the correct position as we can see from its spectrum in Figure \ref{fig:synth_comparison}h. Anyhow, in this result the resolution is lower, in fact, the traces interpolated at the far-offsets include gaps. Moreover, in the first two events, the amplitude is not properly reproduced. \\
The best reconstruction is clearly given by the PINNslope framework. The architecture of the network has the same capacity then the one of the PWD-PINN algorithm and the loss function in fig.\ref{fig:lossessynth}b shows that it has been trained for the same amount of epochs as PWD-PINN with the weight $\lambda=1000$. The key difference in the result is made by the second smaller network that approximates the local slope function. The slope estimate is carried on simultaneously with the interpolation performed by the bigger data network on the original shot gather (Figure 4a), with no filtering required. This simultaneous updating process of data and slope allows for a larger search space to speed up convergence. As can be seen from Figure \ref{fig:synthPINNslope}, the PINN estimated slope closely matches the accurate PWD slope computed from the data of Figure \ref{fig:introsynth}a (ignoring the right-upper part where there are no arrivals and the two approaches clearly extrapolate the values).

\subsection{Field data examples}

\subsubsection{Interpolation beyond aliasing with local slope estimation}
\begin{figure}[h]
    \centering
    \includegraphics[width=1\textwidth]{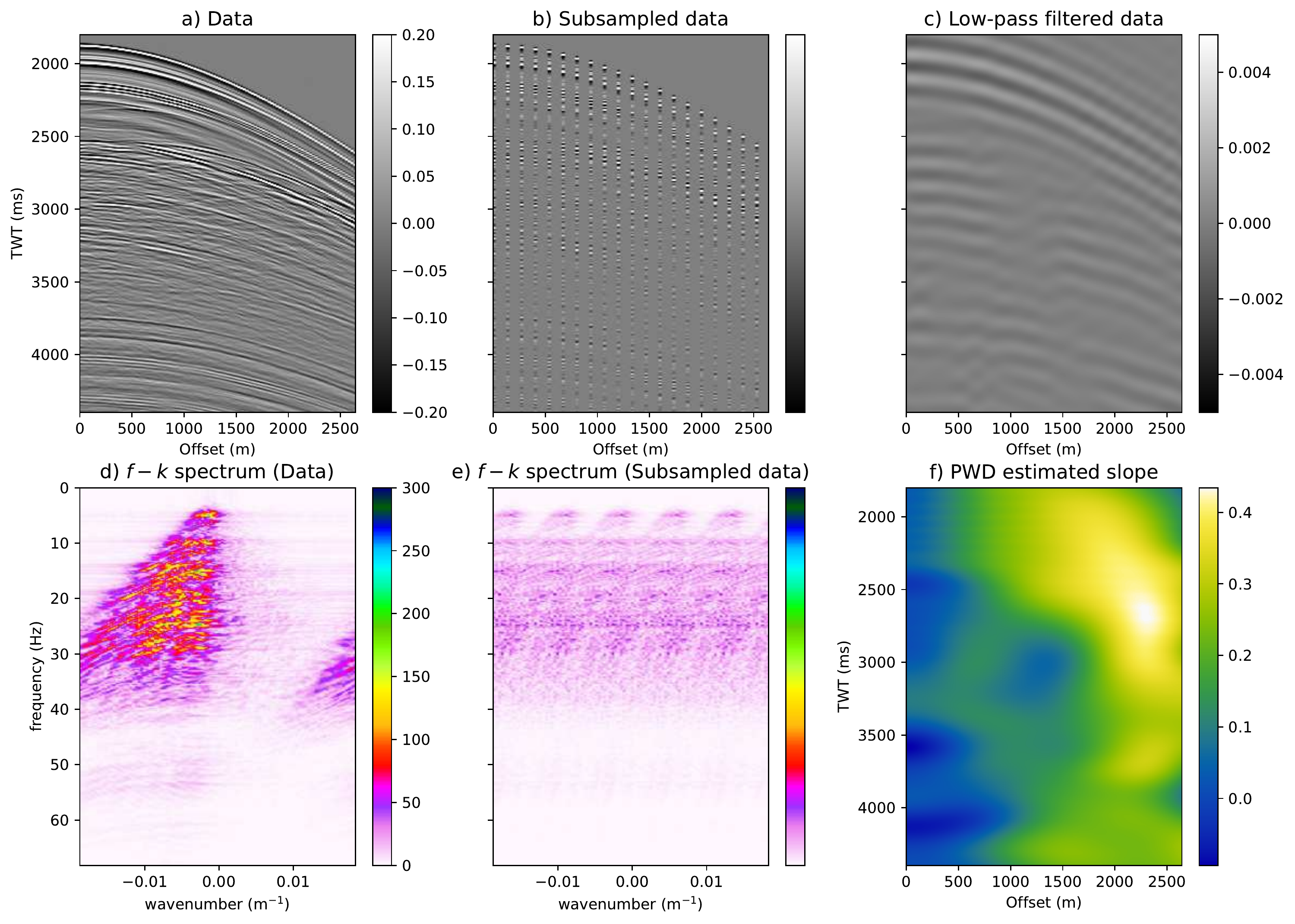}
    \caption{a) The original seismic data, b) subsampled seismic data, c) low frequency data from which the PWD slope has been computed, d) $f-k$ spectrum of the original data, e) $f-k$ spectrum of the subsampled data, f) PWD estimated slope from the low frequency data in Figure \ref{fig:introreal}c.}
    \label{fig:introreal}
\end{figure}

The numerical examples below are performed on a field dataset from the Gulf of Mexico. The trace spacing in the original shot-gather of Figure \ref{fig:introreal}a is 26.7 meters (with a time sampling interval of 0.004 seconds); the data is further subsampled by a factor of 5, increasing the spacing between the traces to 133.5 meters (Figure \ref{fig:introreal}b).

As mentioned earlier, computing directly the slope from the subsampled gather is not feasible and some pre-processing steps are required. Performing the processing steps described above to compute the slope from the low pass data is far more challenging when dealing with the field data in Figure \ref{fig:introreal}e; the part of the signal that is not aliased is very small and does not contain significant energy. The retrieved low frequency data (Figure \ref{fig:introreal}c) are fed into the PWD algorithm and, due to the low frequency nature of the data, the resulting local slope is a low resolution rough estimate of the slope of the high-resolution data (Figure \ref{fig:introreal}f).
The network has an architecture of 4 layers and 512 neurons in each layer, with the number of encoding frequencies set to $N_{X}=8$ and $N_{T}=32$. It is equal to the one used for synthetic data. That is because even if the traces are more complex in the field data, the synthetic traces have an amplitude that does not decrease as much in time. From our initial tests, the network requires the same capacity to easily fit the strong oscillations of the synthetic signals. The slope network (as in the synthetic case) has 2 layers with 2 neurons to estimate a smooth version of the slope field.

\begin{figure}[h]
    \centering
    \includegraphics[scale=0.40]{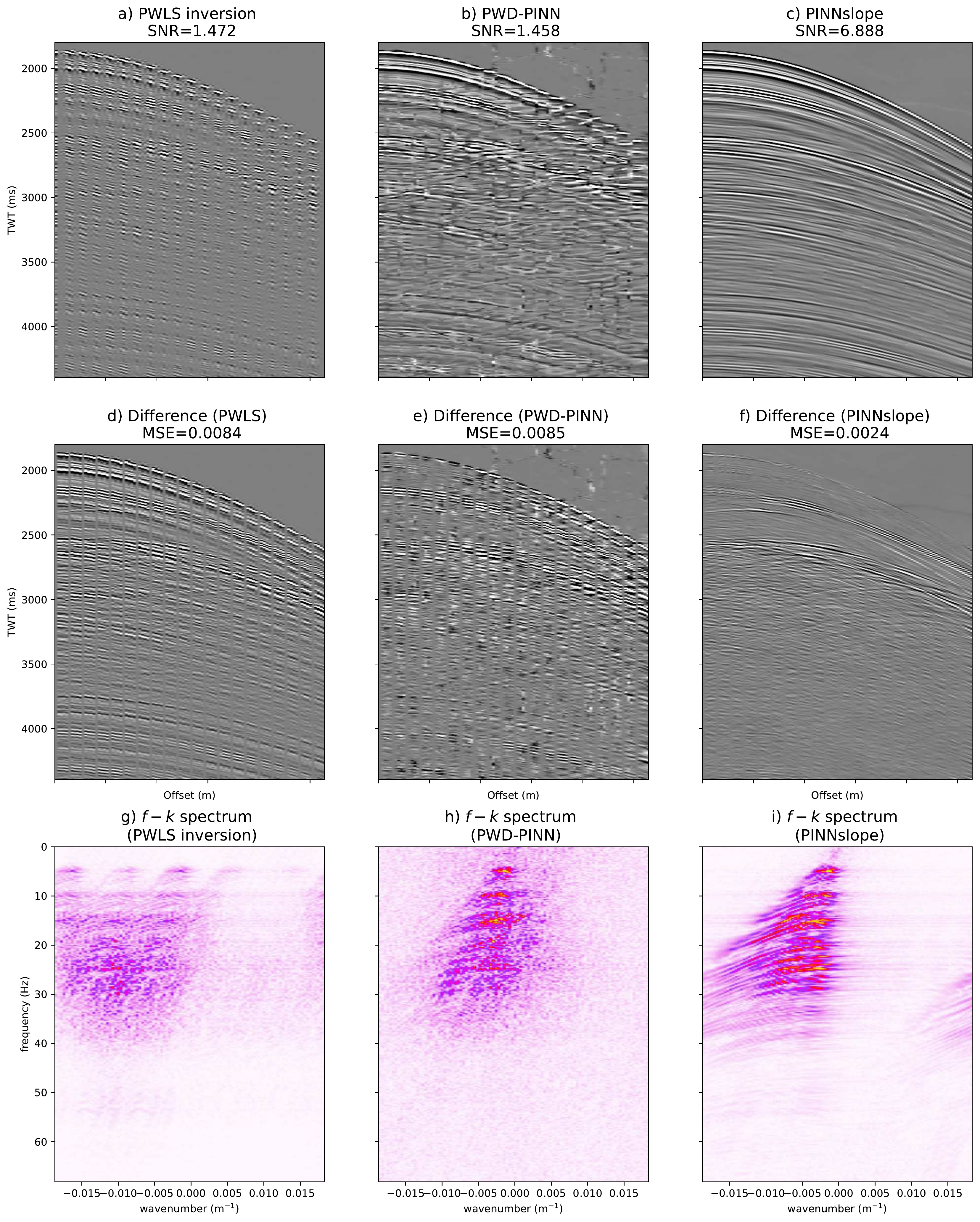}
    \caption{From a) to c): interpolation results of plane-wave regularized least-squares inversion, PWD-PINN and PINNslope. From d) to f) differences between the original data and PWLS inversion, PWD-PINN, and PINNslope. From g) to i) $f-k$ spectrum of PWLS inversion, PWD-PINN and PINNslope. The SNR and MSE values are provided on top of the panels.}
    \label{fig:realresults}
\end{figure}

Figure \ref{fig:realresults}a displays the reconstruction produced by the PWLS algorithm; in this case, the result is not comparable with that obtained in the synthetic data example in terms of reconstruction quality. Given the noise present in the field data and a poor slope estimate (when compared to the synthetic one), the PWLS algorithm cannot reconstruct the missing component of the data.

\begin{figure}[h!]
    \centering
    \includegraphics[width=1\textwidth]{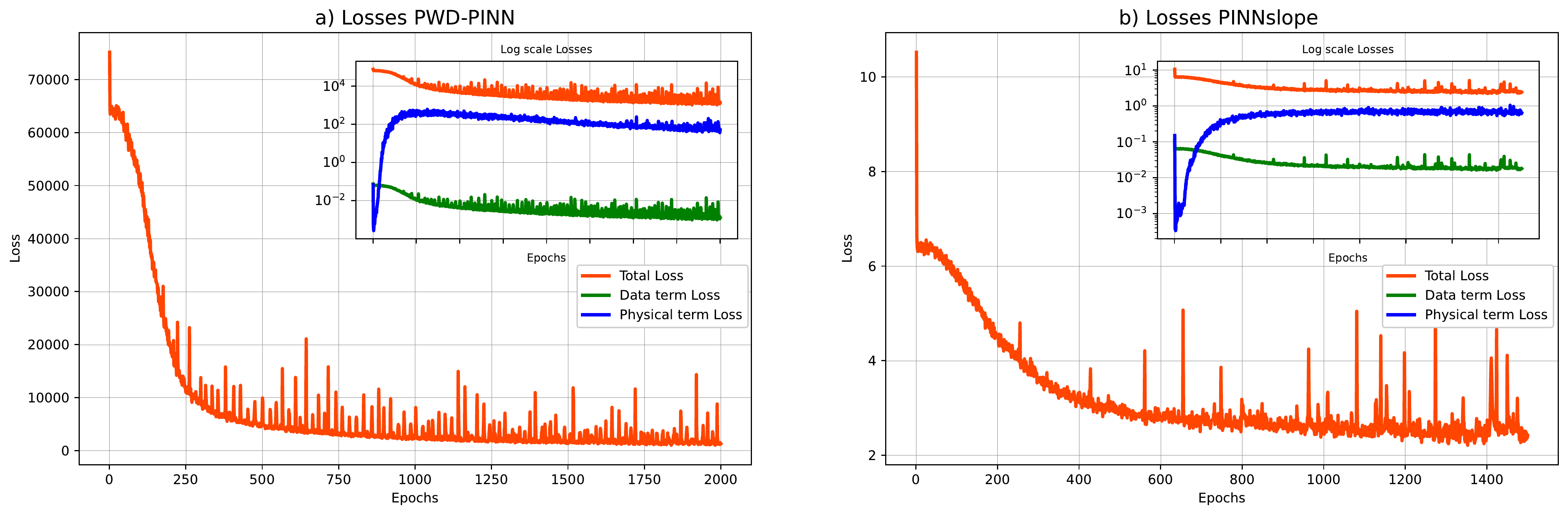}
    \caption{a) Loss curves of the PWD-PINN training on real data, b) loss curves of the PINNslope training on real data. In the small box inside the two plots, we show the contribution of each term of the loss function in the log scale, that is: data-term (in green), physical-term (in blue), and total loss (in orange).}
    \label{fig:lossesreal}
\end{figure}

The PDW-PINN algorithm (Figure \ref{fig:realresults}b) does not achieve a good result as well.  Only the interpolation in the near-offset could be considered reasonable. In the far-offsets, the reconstruction is worse as we have already observed in the synthetic data. In this part of the dataset, the PWD estimated slope is prone to errors and it does not allow a good interpolation. The loss curves (Figure \ref{fig:lossesreal}a) show that we had to increase the $\lambda$ parameter to very high values ($\lambda=10000$) to make the network properly fit the traces. The network struggles to accurately fit the traces if the accuracy of the local slope is poor, as this will negatively affect the PDE term contribution in the loss function.\\
In contrast, the PINNslope approach admits a good performance (with the parameter $\lambda=100$); most of the energy has been restored and the aliasing has been successfully suppressed. The extra degrees of freedom provided by the small slope network helped the convergence.

\begin{figure}[h]
    \centering
    \includegraphics[scale=0.4]{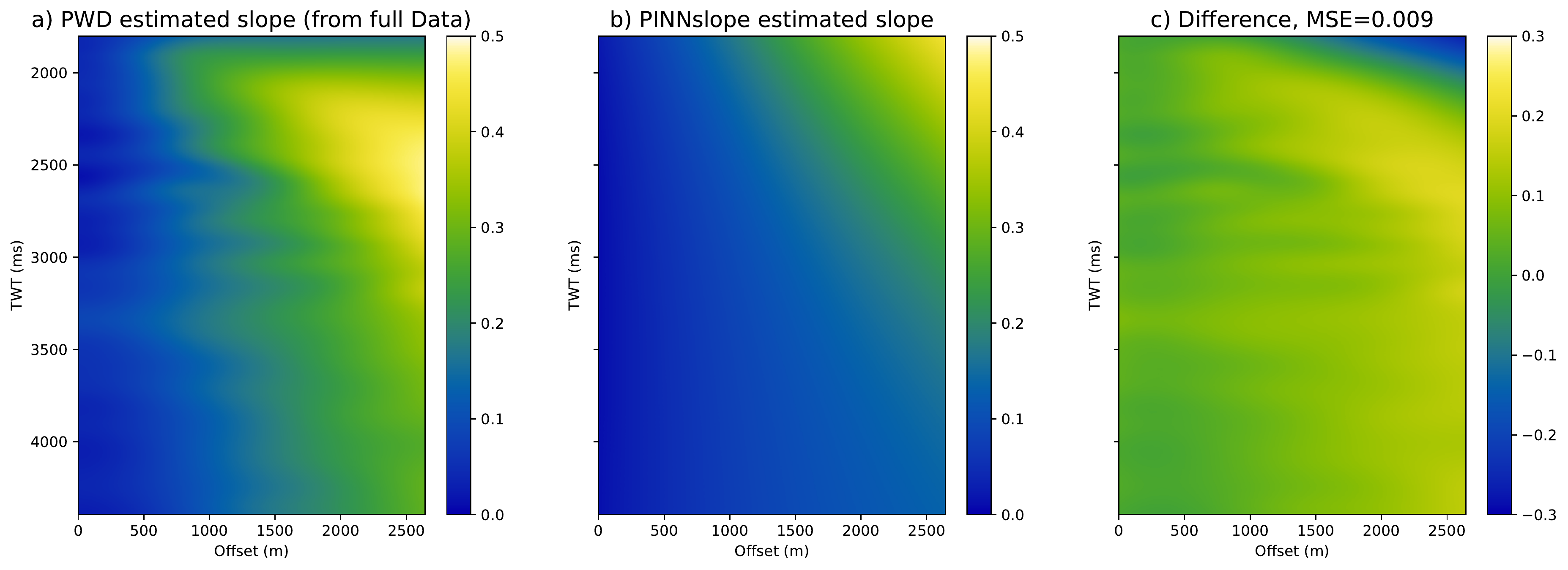}
    \caption{a) The slope estimated with the PWD algorithm from the data shown in Figure \ref{fig:introreal}a, b) the slope estimate through the PINNslope framework while simultaneously interpolating the data in Figure \ref{fig:realresults}e, c) the difference between Figure \ref{fig:realPINNslope}a and Figure \ref{fig:realPINNslope}b. }
    \label{fig:realPINNslope}
\end{figure}

In Figure \ref{fig:realPINNslope}, the PINN estimated slope is compared to the PWD slope computed from the full data (Figure \ref{fig:introreal}a). The PINN slope is smoother than the PWD one and again admits generally lower values, probably because it has been estimated on less dense data. However, the overall trend of the PINNslope slope is correct and its smoothness serves its purpose in the plane-wave regularization term. If compared to the PWD slope in Figure \ref{fig:introreal}f,  which is the realistically achievable slope when we try to solve an interpolation problem of this kind, the one from PINNslope is a better and more precise estimate.

The residuals shown in Figure \ref{fig:realresults}f and Figure \ref{fig:realresults}e, differently from the synthetic data, are partially due to the field data containing secondary events with conflicting dips that cannot be recovered by our method. We note that this is a general weakness of interpolation methods relying on the plane-wave PDE.

\subsubsection{Performance assessment}

In this section, the PINNslope framework is tested on a harder interpolation task where fewer traces are available. The aim is to assess its performance on the current dataset and evaluate its limitations. Moreover, as subsequent shot gathers in the dataset will have only minor changes between them, we test the convergence behavior of the pre-trained network when applied to the next gathers in the dataset.
We first apply PINNslope on the shot-gather subsampled by various factors: 6, 7, and 8 (respectively, 160.2 meters, 186.9 meters, and 213.6 meters intervals between each trace).
Finally, we also test the ability of our framework to interpolate a dataset with a large gap of traces (i.e. 15 traces, for a total of 400.5 meters gap) positioned in the middle of the gather. This is a challenging task for the network, which has to rely solely on the information obtainable from the left and right side of the gather, as in the gap region there is no knowledge of the shape of the arrivals nor of their slope field. So far, we are not aware of any interpolation algorithm that can solve this category of interpolation tasks in an automatic and physically driven manner.\\

\begin{figure}[h!]
    \centering
    \includegraphics[scale=0.36]{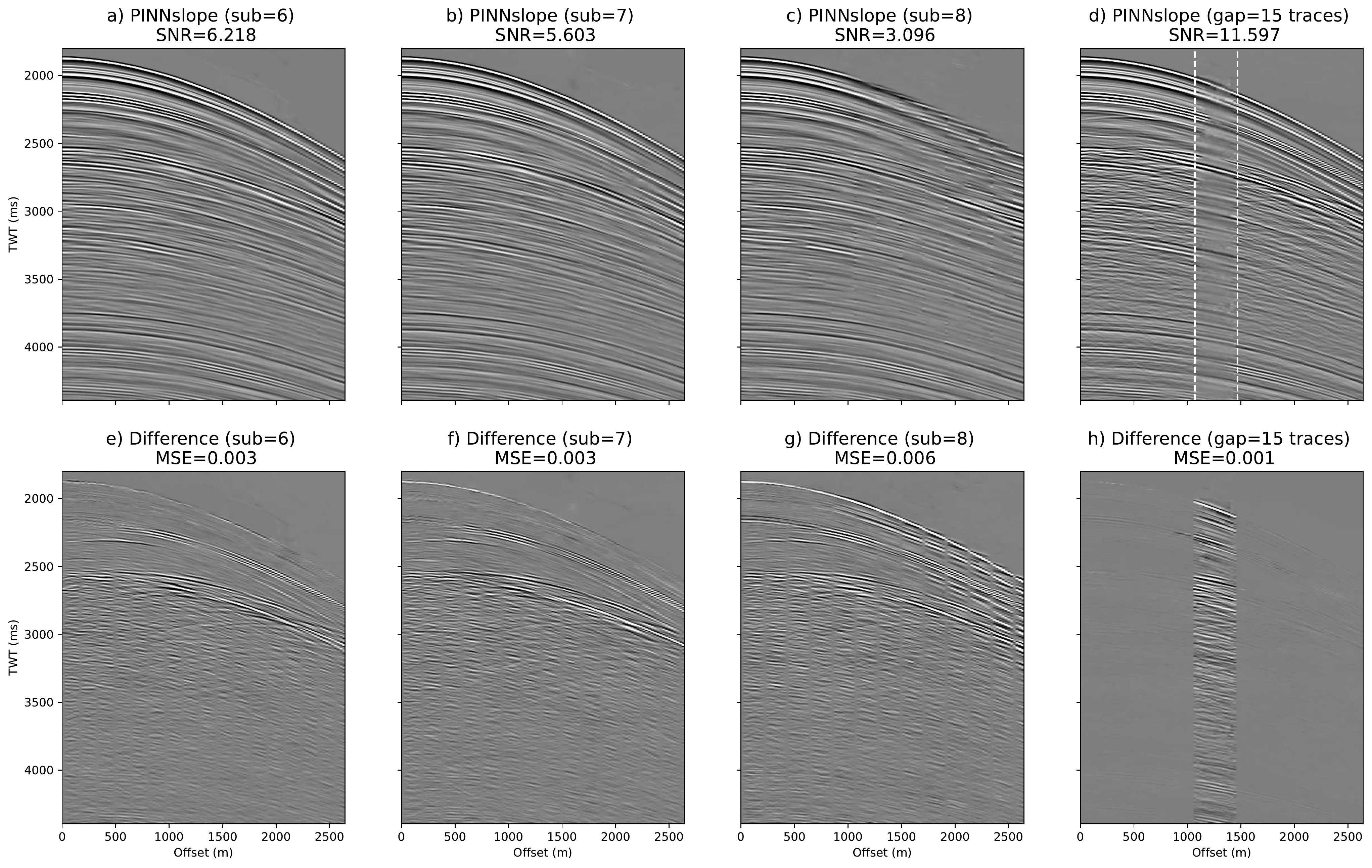}
    \caption{The performance of PINNslope with high subsampling rates and a particular case where an entire part of the gather is missing. From a) to d): PINNslope results obtained starting from data subsampled by a factor of 6, 7, 8 and from data that contain a large gap of 15 traces (white hatched lines delimit the gap area). From e) to h): difference between the original data and the PINNslope reconstructions. The MSE and SNR values are provided
    on top of the panels.}
    \label{fig:perf}
\end{figure}

The performance results are shown in Figure \ref{fig:perf}. For this shot gather, a higher subsampling of 6 and 7 (respectively Figure \ref{fig:perf}a and Figure \ref{fig:perf}b) does not impact the network performance. The signal-to-noise ratio is maintained almost constant from the result described in the previous section and the arrivals are satisfactorily interpolated. The framework starts to face some challenges when the gather is subsampled by a factor of 8 (213.6 meters interval between the traces). In the near-offset, the interpolation is still accurate but as the dip of the arrival starts to increase, PINNslope is unable to retrieve the correct slopes and struggles in interpolating the arrivals. We consider this sampling to be the framework threshold limit for this shot gather (and the frequency range involved). The presented results are obtained by increasing the number of epochs as the subsampling increases (1500 epochs for a subsampling of 6, and 2500 for a subsampling of 8). In the case of the higher subsampling factor, it was additionally needed a $\lambda$ value of 10000 instead of the usual 100.\\

\begin{figure}[h!]
    \centering
    \includegraphics[scale=0.36]{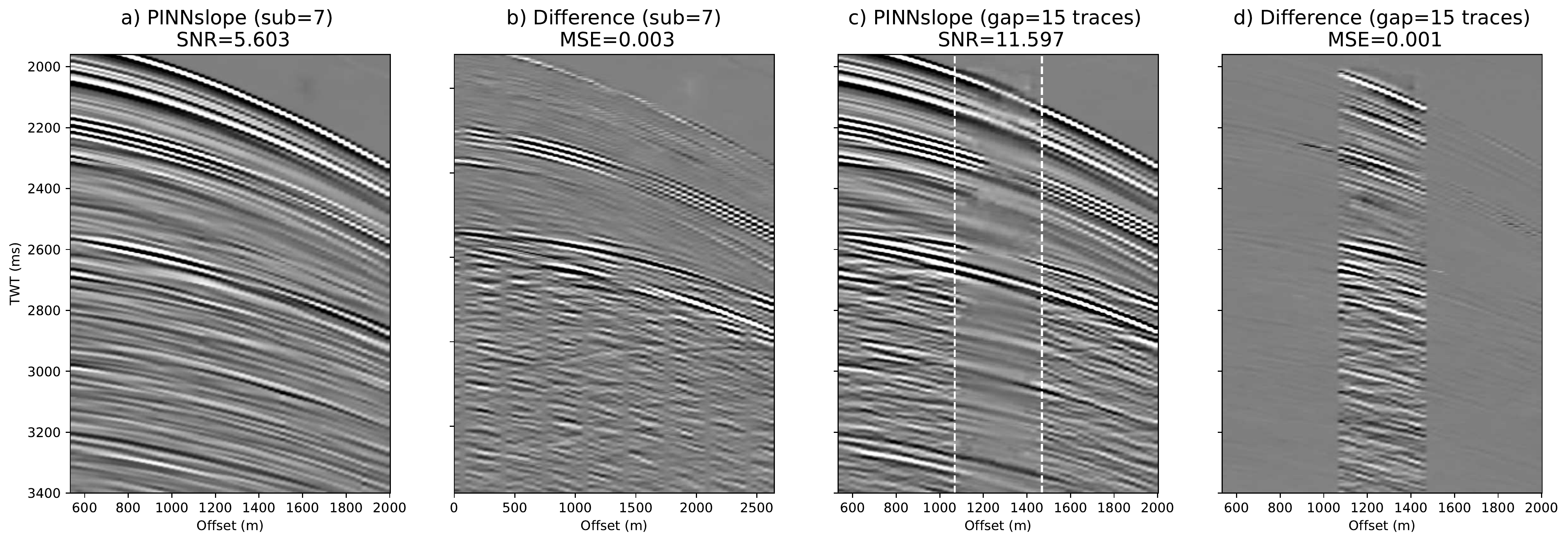}
    \caption{Zoom in of some of the results in Figure \ref{fig:perf}. a) PINNslope interpolation of data subsampled by a factor of 7, b) difference between the original data and \ref{fig:perf_zoom}a, c) PINNslope interpolation result of data containing a large gap (white hatched lines delimit the gap area), d) difference of original data and \ref{fig:perf_zoom}d. The SNR and MSE values are provided on top of the panels.}
    \label{fig:perf_zoom}
\end{figure}

\begin{figure}[h!]
    \centering
    \includegraphics[scale=0.40]{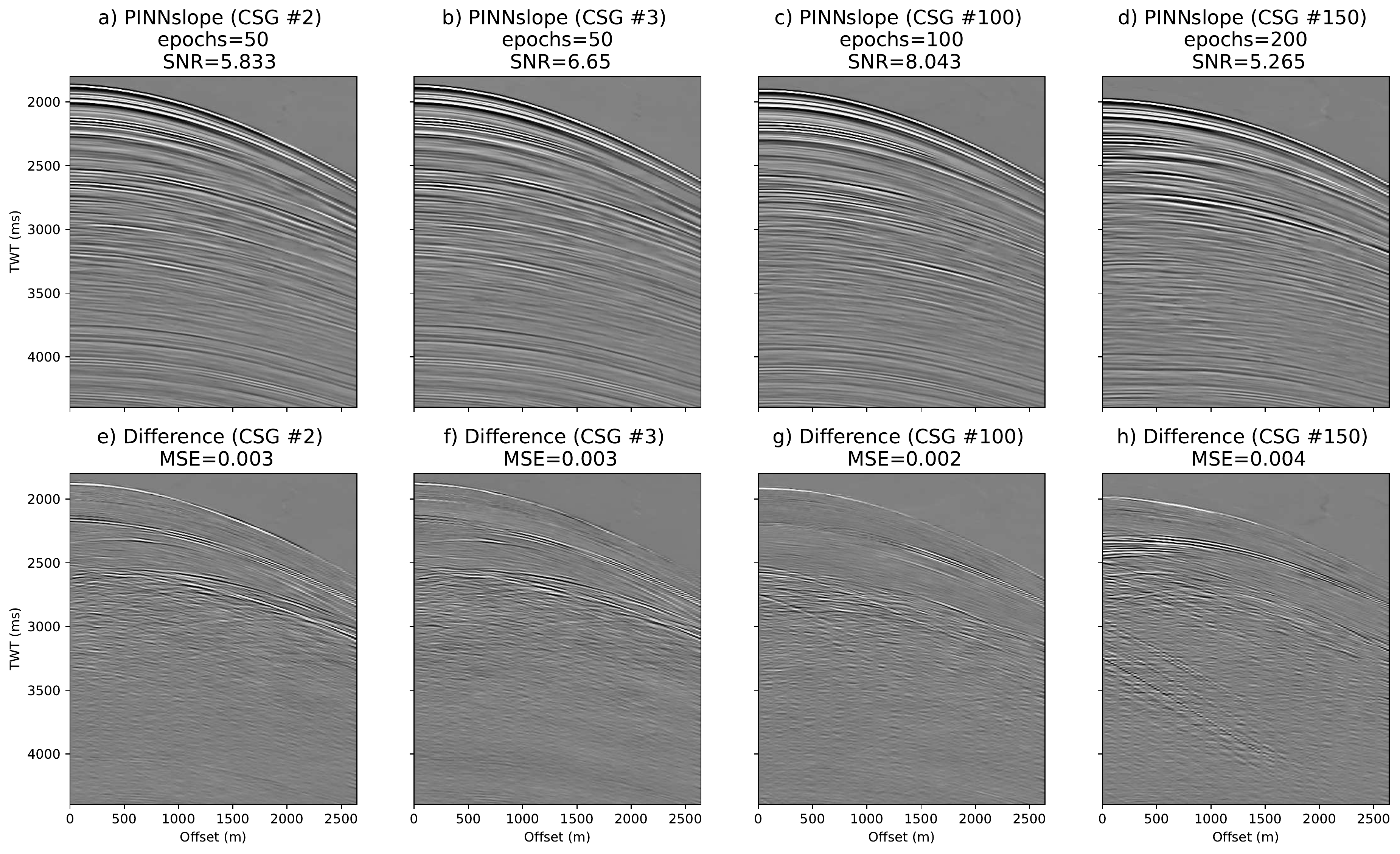}
    \caption{The results of fine-tuning PINNslope on different gathers from the one used to pre-train the network (Figure \ref{fig:realresults}a). From a) to d): fine-tuned PINNslope interpolation result of the gathers at the specified position of the original dataset. From e) to h): differences between the original gather at the specified position and the corresponding PINNslope reconstruction. The MSE and SNR values are provided on top of the panels. }
    \label{fig:perf_csg}
\end{figure}

By enlarging Figure \ref{fig:perf_zoom} at early times we can see the main difference between the original data and the PINNslope interpolated data. Most of the events are well reproduced. However, as the energy corresponding to the often weaker events with conflicting dips is missing, the interpolated data looks cleaner than the original one. As mentioned before most algorithms that rely on slope estimation to perform their tasks cannot leverage on the energy of the second order (conflicting) dips for its reconstruction, unless some additional slopes are included in the process.\\
The result of the gap interpolation (is shown in Figure \ref{fig:perf}d and Figure \ref{fig:perf_zoom}c). The PINNslope framework is able to extrapolate the main reflections from the left and right sides of the gather and connect them together. Of course, as expected significant errors arise in the middle of the gap, as the algorithm lacks any type of information about the missing region.
Is it worth mentioning that this experiment has been performed to evaluate the interpolation capabilities of neural networks and especially of PINNs. Figure \ref{fig:perf_csg} shows the convergence capabilities of the PINNslope network (pre-trained on the gather of Figure \ref{fig:realresults}a with a subsampling factor of 5) on the subsequent gathers of the dataset (that have been subsampled also by a factor of 5).

Specifically, the interpolated gathers of Figure \ref{fig:perf_csg}a and Figure \ref{fig:perf_csg}b are consecutive to the one used as the benchmark until now (respectively, position 2 and 3 of the dataset, while the benchmark gather is the first one in the dataset). Instead, the ones in Figure \ref{fig:perf_csg}c and Figure \ref{fig:perf_csg}d are way more distant (respectively, position 100 and 150 of the dataset). The idea is to use only one pre-training step (on the gather of Figure \ref{fig:realresults}a) and a small fine-tuning step to make the network fit the subsequent gather. As we can see, the number of epochs needed for fine-tuning the network depends on how different the gathers are from the one used for the pre-training. Although, all the gathers have been well reproduced with a small number of epochs (from 50 to 200 epochs depending on the shot gather position, see titles of Figure \ref{fig:perf_csg}a, b, c, d) compared to the full training.\\
Finally, we demonstrate the mesh-free property of PINNslope (and PINNs in general) by evaluating the network at different grid points than those it has been trained on.

\begin{figure}[h!]
    \centering
    \includegraphics[scale=0.65]{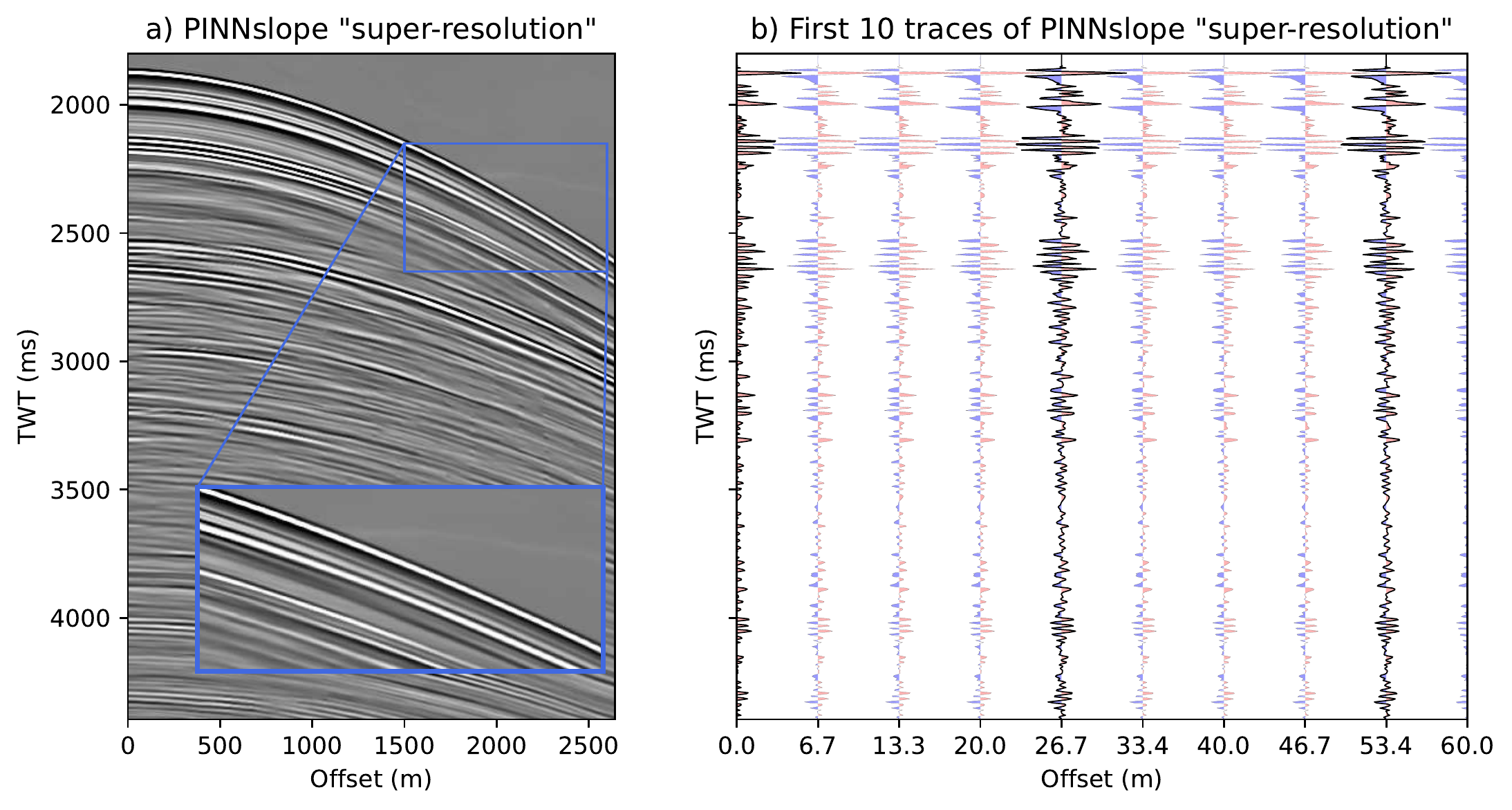}
    \caption{Result of PINNslope evaluated on a denser grid. a) super-resolved interpolated shot-gather, b) plot of the first ten traces from the gather in a), with highlighted in black the traces evaluated at the training grid points locations, and in gray the traces evaluated at the additional points of the newly denser grid.}
    \label{fig:s_res}
\end{figure}

Figure \ref{fig:s_res} shows the result of the PINNslope interpolation (starting from the data subsampled from a factor of 5 of Figure \ref{fig:introreal}b). The PINNslope training has been performed by feeding to the network random batches of grid points sampled from a coordinate grid with a trace interval of 26.7 meters and a time sample interval of 0.004 seconds (which are the values of the coordinate grid utilized for the training and evaluation of the benchmark dataset employed in the previous experiments of the section "Field data examples"). In this example, PINNslope has been evaluated on a denser grid with respect to the one used in training. The denser grid has a trace spacing of 6.7 meters while the time sample has been maintained at 0.004 seconds. This result reveals that PINNslope can provide 'super-resolution' interpolation capabilities, as shown from the zoom-in plot on the first reflections of Figure \ref{fig:s_res}a. In the left plot (Figure \ref{fig:s_res}b), the first ten traces of the shot gather are displayed. The bold black traces correspond to the ones positioned at the original training interval of 26.7 meters, while the gray traces between the bold ones correspond to the network prediction at the denser grid points.

\begin{figure}[h!]
    \centering
    \includegraphics[scale=0.65]{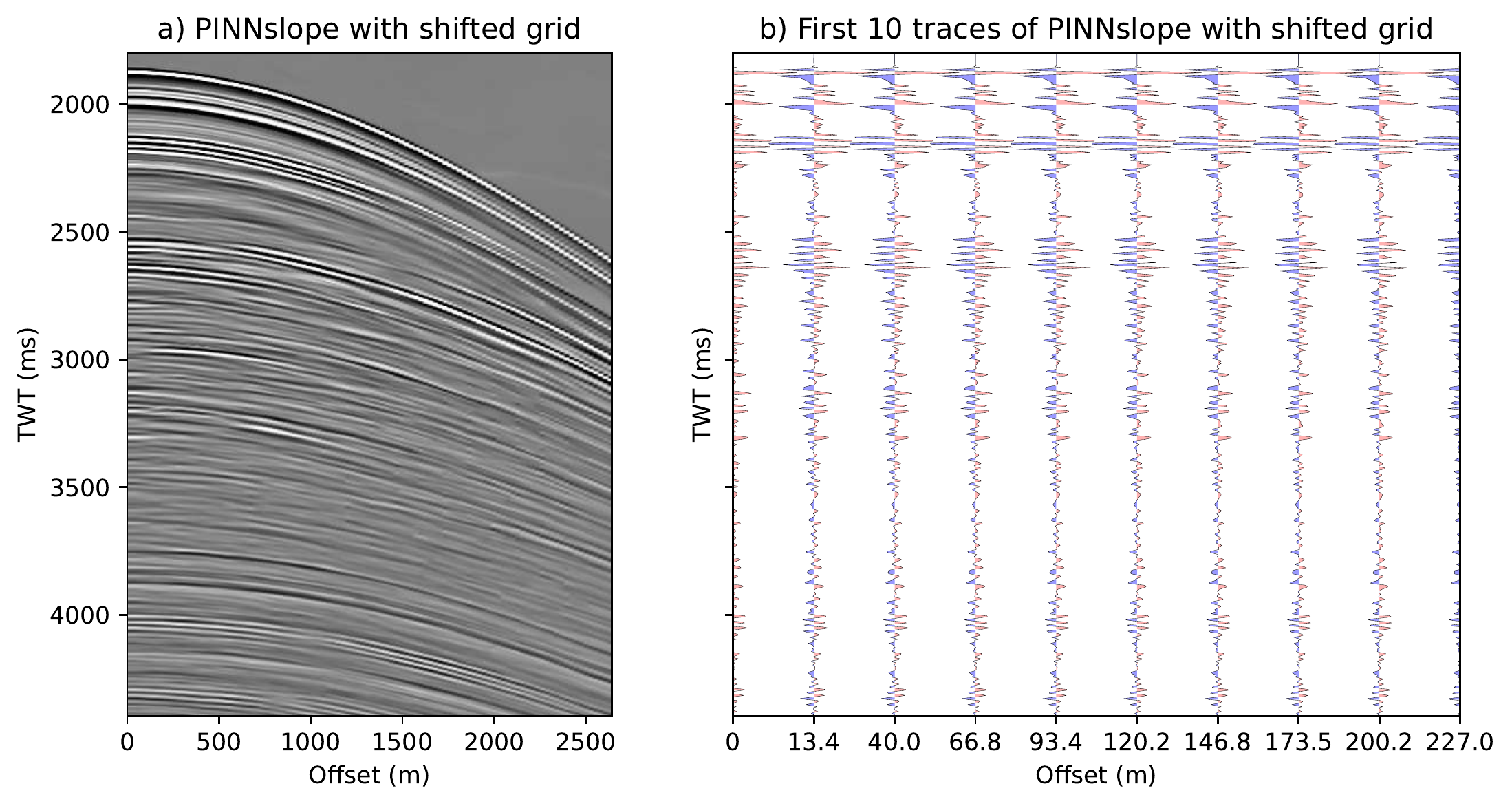}
    \caption{Result of PINNslope evaluated on a shifted grid with respect to the one used in training. a) PINNslope interpolation result on the newly shifted grid, b) first ten traces from the shot-gather in a), showing the traces evaluated at the new offset coordinate positions.}
    \label{fig:shift_g}
\end{figure}

Figure \ref{fig:shift_g} presents a slightly different example. PINNslope has been trained as before, starting from a gather subsampled by a factor of 5 on a grid with the same features described above (trace spacing of 26.7 and time samples interval of 0.004 seconds). In this case, the network has been evaluated on a grid with the same trace spacing of 26.7 meters but the coordinate points have been shifted by half a trace interval with respect to the coordinate positions of the training grid points. Such a need may arise in time-lapse applications when one is interested in defining a common grid for multiple datasets (e.g., transferring the monitor dataset onto the grid of the baseline dataset). As we can reckon from these two experiments, there is no need to train a second time the network if the coordinates or the grid points interval change because the results obtained by training and evaluating on the same grid can be perfectly reproduced by simply applying the trained network on a different coordinate grid. 

%% file: Sections/Discussions.tex
\section{Discussion}

Out of all the slope-assisted interpolation methods discussed in this paper, PINNslope achieves the best performance. It can interpolate the missing data eliminating the aliasing present in the data whether they are related to coarse recordings or the presence of obstacles during the acquisition. Parameterizing the plane-wave PDE in the loss function with a small neural network does not significantly affect the runtime with respect to PWD-PINN. For PWD-PINN and PWLS inversion a number of pre-processing steps should be performed before being able to compute the local slope to be utilized in the PDE regularization term. This is also a time consuming part of these approaches. Although, the PWLS inversion is almost instantaneous compared to the PINNs. \\
All three methodologies present similar drawbacks due to the plane-wave approximation, which does not allow us to reconstruct events with conflicting dips present in the gathers. In fact, the slopes estimated by the PWD filters (or any other algorithm that can estimate slopes) and PINNslope are computed with respect to the main events and they are not able to retrieve information for the conflicting slopes (slopes that have opposite or different direction) in the data. The PINNslope framework cannot reproduce the events that have a reverse slope with respect to the main trend of the arrivals. Moreover, the estimate of the slope on a data point where two events are crossing each other will always induce errors in the data fit. Currently, it is not possible to compute the value of two slopes on one single point of the dataset unless we include two slopes in the framework, which will be investigated in the future.\\
PINNslope has the potential to fill a large gap of traces via a physically driven approach not achievable by any other type of algorithm, furthermore estimating the full slope field. This example illustrates the interpolation capabilities of this type of implementation. A substantial help to the network fitting ability is derived from the \textit{positional encoding} layer. In this study, we considered uninformative to display the interpolation result of PINNslope without \textit{positional encoding}, as the resulting shot gather amplitude would be close to zero even when utilizing a network with a higher capacity. In past experiments and implementations, various methodologies have been tested to make the network able to fit complex signals as field seismic traces. \textit{Locally-adaptive activation functions} \cite{Jagtap2020} have been implemented along with a \textit{Sin()} activation function \cite{Brandolin2022}, to allow the network to be more expressive without the need to extend its capacity, but still it was insufficient. In addition, a technique named frequency upscaling by the mean of neuron splitting \cite{Huang2022} has been tested achieving good results in fitting complex high frequency signals. One of its drawbacks is that it injects low level noise into the reconstructed result and requires to train various time the network to adequately fit the entire frequency content of the dataset. On the other hand, \textit{positional encoding} solved this issue allowing the network to accurately reproduce the field data and achieve a faster convergence, without the need to train several times on the same dataset (as frequency upscaling with neuron splitting).    

%% file: Sections/Conclusions.tex
\section{Conclusions}
We introduced a novel PINN framework in the seismic signal processing field for simultaneous seismic data interpolation and local slope estimation.
It is discernible also as an innovative procedure for local slope attribute estimation of complex subsurface images.
The results obtained with PINNslope are compared on synthetics and field datasets against PWD-PINN and PWLS, to better examine its performances against two approaches that rely on pre-computed slopes to perform the interpolation. We found that introducing a second network to estimate the local slope attribute while at the same time interpolating the aliased data achieves better results in terms of signal-to-noise ratio, while also improving the overall network convergence. The positional encoding layer was a fundamental addition to the architecture and helped overcome previous difficulties such as high frequency fitting and noise introduction during the interpolation process.
The PINN estimated slopes look accurate and consistent with the interpolated data, their accuracy is comparable to the one obtainable with the plane-wave destruction filters estimate.

%% file: Sections/Acknowledgement.tex
\section*{Acknowledgments}
This publication is based on work supported by the King Abdullah University of Science and Technology (KAUST). The authors thank the DeepWave sponsors for supporting this research.